*chemosensors*  MDPI*Article*

# Bioinspired Materials for Sensor and Clinical Applications: Two Case Studies

Eleonora Alfinito [1,*], Mariangela Ciccarese [2], Giuseppe Maruccio [1,3], Anna Grazia Monteduro [1,3] and Silvia Rizzato[1,3]

[1] Department of Mathematics and Physics 'Ennio De Giorgi', University of Salento, I-73100 Lecce, Italy
[2] Oncology Unit "Vito Fazzi" Hospital-Lecce, I- 73100 Lecce, Italy
[3] Omnics Research Group, CNR-Institute of Nanotechnology, INFN Sezione di Lecce, Via per Monteroni, I-73100 Lecce, Italy

* Correspondence: eleonora.alfinito@unisalento.it**Abstract:** The growing interest in bio-inspired materials is driven by the need for increasingly targeted and efficient devices that also have a low ecological impact. These devices often use specially developed materials (e.g., polymers, aptamers, monoclonal antibodies) capable of carrying out the process of recognizing and capturing a specific target in a similar way to biomaterials of natural origin. In this article, we present two case studies, in which the target is a biomolecule of medical interest, in particular, $\alpha$-thrombin and cytokine IL-6. In these examples, different biomaterials are compared to establish, with a theoretical-computational procedure known as proteotronics, which of them has the greatest potential for use in a biodevice.

**Keywords:** biosensors; aptamer; antibody; MIPs**Citation:** Alfinito, E.; Ciccarese, M.; Maruccio, G.; Monteduro, A.G.; Rizzato, S. Bioinspired Materials for Sensor and Clinical Applications: Two Case Studies. *Chemosensors* **2023**, *10*, x. https://doi.org/10.3390/xxxxx

Academic Editors: Pi-Guey Su and Camelia Bala

Received: 29 December 2022
Revised: 15 February 2023
Accepted: 14 March 2023
Published: date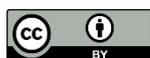

**Copyright:** © 2023 by the authors. Submitted for possible open access publication under the terms and conditions of the Creative Commons Attribution (CC BY) license (https://creativecommons.org/licenses/by/4.0/).## 1. Introduction

Interdisciplinarity and cross-fertilization offer relevant opportunities for the advancement of knowledge at the border between different disciplines.

This is achieved by sharing objectives, techniques, and above all, materials. Refined by evolution to carry out specific actions with high efficiency, biological matter (especially macromolecules) shows extreme functionality and a level of specialization that make it particularly interesting for many applications in the biomedical and technological fields. Applications range from biosensing and diagnostics [1] to drug development [2–4] and even molecular electronics [5,6]. Furthermore, this type of matter is an inspiration for the design of synthetic biomimetic materials that they can imitate, and they can often improve on its specificities. Progress in this field requires a rethinking of both the experimental and theoretical methods of investigation. The nanoscale realm of macromolecules poses formidable challenges. Experimentally, traditional optical microscopy fails in addressing matter at this scale, and indeed, the field benefited from progress in nanoscience and nanotechnology and the advent of novel approaches such as electron, scanning probe, and confocal microscopies with associated spectroscopies [7,8]. This allowed for investigations with unprecedented level of details and insight, probing the structure, conformation, function, folding/unfolding, and interactions of biomolecular building blocks down to a single-molecule regime.

Furthermore, at the nanoscale, classical and quantum effects converge, and because of this, many standard theoretical approaches also fail. The First principle methods [9], mainly based on a quantum analysis, still limit the investigation to small regions of a biomolecule, due to computational limitations. On the other hand, macroscopic approaches

*Chemosensors* **2022**, *10*, x. https://doi.org/10.3390/xxxxx    www.mdpi.com/journal/chemosensors

mainly focus on classical phenomena. Therefore, to enable classical and quantum interconnections and to address the macromolecular scale, new methods are much sought after. In this regard, proteotronics [10] have been developed as a new investigative procedure that maps the macromolecule into a network of interactions. Proteotronics is an interdisciplinary technique that uses the crystallographic structure of macromolecules to produce a graph. By introducing suitable physical interactions between the nodes of the network, it transforms the graph into an impedance network that it solves with numerical methods. The strategy is, therefore, a bottom-up process that reconstructs the global response of a distributed impedance system (the macromolecule), starting with its elements (amino acids/nucleobases). The applications include: the analysis of the structure of single macromolecules or molecular complexes [11–13]; the prediction of the most stable molecular complexes and interactions (relevant for pharmaceutical research) [14–17]; the evaluation of some specific processes, including charge transfer and its influence on the expected electrical responses in bioelectronic devices [10,17].

## 2. Materials and Methods

### 2.1. Materials

Nature has selected antibodies (Abs) as the best weapons against microbial attacks, identifying the intruder with high accuracy, thus allowing the immune system to fire it. Monoclonal antibodies (mAbs), designed to bind to selected targets, partially reproduce natural antibody behaviour [18,19]. mAbs are currently used in medicine for the treatment of various types of diseases, from tumors to hypercholesterolemia; furthermore, their utilization in biosensors is increasing. On the other hand, mAbs show critical issues on several aspects, for example, their own production, which is costly and time consuming [18], the propensity to produce side effects [18,19], and, not secondarily, the related ethical issues [20].

To overcome these limits, new materials capable of mimicking the performances of antibodies are continuously developed, including, for instance, antibody derivatives such as antigen-binding fragments, Fabs, or aptamers and polymeric matrices (molecularly imprinted polymers (MIP)).

The antibody fragments allow for a non-mammalian expression system, i.e., they can be expressed even in lower organisms, and this reduces costs and ethical concerns and speeds up production times. On the other hand, no reduction of specificity for targets is observed [18]. Concerning aptamers, they are often cited as "synthetic antibodies", which is an attractive, but not entirely correct, definition because they are not proteins that are made up of amino acids, but single strands of DNA or RNA. They are produced artificially with the SELEX technique, introduced by Tuerk and Gold in 1990 [21], which, in principle, allows us to select aptamers for any type of target. They are very appreciated for the high biocompatibility, which does not produce relevant side effects, as well as for the high affinity to the target and the ease of modification [22]. Concerning the cons, their production is, again, quite costly and time consuming; furthermore, they are easily degradated, and this limits their clinical applications [14].

In general, the fabrication of MIPs consists of the polymerization of the functional monomer(s) in the presence of template molecules, in order to allow for the formation of specific recognition sites after the removal of the templates. Usually, bulk polymerization is the strategy employed in MIP synthesis for small molecule models, but it does not work well with large macromolecules, e.g., G-proteins, as their removal/binding from/to the buried MIP cavity is difficult, resulting in low sensitivity and long response times. Several strategies have been introduced to enable both macro and small molecule imprinting, such as the electropolymerization approach, which allows for easy modification of a conventional electrode and facilitates the diffusion of macrobiomolecules within the polymer matrix and their rebinding, or the use of nano- or microparticles as an additional template to fabricate macroporous MIPs with molecular cavities located on the surface of the pores [23–28].

At present, while monoclonal antibodies, antibody derivatives, and aptamers are developed for both clinical and technological applications, MIPs are mainly employed in technological applications, such as biosensor development or for remediation purposes [29–32].

In all the indicated applications, the material (hereafter referred to as the receptor) captures a specific target (ligand), and, in doing so, it implements a specific binding mechanism. Knowing the type of binding mechanism used by the receptor is critical to understanding, and possibly controlling, receptor–ligand affinity.

In both experiments and modelling, the question of which scenario is the best suited to describe the dynamics of binding is a long-debated issue [33]. The first model to be envisaged was the lock and key, which assumed the existence of the receptor in a single form, perfectly complementary to the ligand. Subsequently, the role of conformational change in the binding mechanism has been confirmed in several receptor–ligand complexes, especially when small molecules, such as aptamers, are involved [12,13]. Therefore, alternate scenarios were formulated, which describe the conformational change as a transition among different semi-equilibrium states [34]: the induced fit describes a reciprocal adaptation of the cognate biomolecules, in which both the biomolecules land to an energy minimum, while the conformational selection proposes the coexistence of biomolecules in different conformations, i.e., energy states, and the binding affects only those having the appropriate conformation.

These three mechanisms are probably the borderline cases of more complex situations, in which all 3 coexist. On the other hand, while for aptamers and (natural/synthetic) antibodies, a reciprocal structure adaptation (conformational selection/induced fit) is the dominant binding mechanism, and MIPs capture ligands mainly using an optimal interlocking mechanism (lock and key). Aptamers and antibodies perform a conformational change that goes with an energy state transfer, and MIPs bind to ligands in their energy minimum state (or the fraction of the sample that is in that state), thus inducing no conformational changes. Lacking the adaptation, the dynamic of recognition is, in a certain sense, passive and works very well with inorganic ligands [31], offering the advantage of being accurate and providing highly selective results, as well as being fast and cheap, compared to conventional clinically applied methods for biomedical markers, pathogens, and toxin detection [35–37], such as enzyme-linked immunosorbent assay (ELISA) and real-time polymerase chain reaction (RT-PCR) [36].

The difference in the binding mechanism is probably at root of the different performances observed between MIP-based and aptamer-based devices [32]. In particular, aptasensors show higher sensitivity, with respect to MIP-devices and faster reaction, while they are less prominent in selectivity and stability [32]: specifically, MIPs are able to detect, with extreme selectivity, ligands in the native state, and this binding should be stable if the polymeric matrix does not degrade. Conversely, a device using aptamers, should be able to also capture only fractions of the ligand, as well as ligands in different conformations [12,13], and this can explain the higher sensitivity and the lower selectivity; finally, given the flexibility of the receptors, the bond has a limited duration.

In the following, we will focus on aptamers and antibody derivatives, analyzing different receptor–ligand complexes, in terms of *topological complementarity*: it is due to the dynamic binding mechanisms they implement and, as such, depends on the specific environmental conditions [17].

2.1.1. Cytokine IL-6 Receptors

Cytokine IL-6 belongs to a protein family activated through the glycoprotein gp-130. In particular, it forms a heterotrimer with the IL-6R receptor, at the site1, S1, and binding site, and in an almost antithetical way, it binds gp-130 at site 2, S2. Finally, the formation of the quaternary hexameric structure, composed of two identical trimers, involves a third site on IL-6, S3 [38,39] (See Figure 1). IL-6 has a central role in both physiological and pathological processes, and it is elevated in many solid tumors, including breast cancer [40]. It has been demonstrated that deregulation of IL-6 signaling pathway is a trigger point in proliferation, migration, and adhesion among tumors [41]. High levels of IL-6 in breast cancer tissues induce Jagged-1 expression to promote cell growth and maintain the aggressive phenotype [42]. Due to the complexity of the structure and the interest of this protein in various inflammatory mechanisms, various biomolecules able to bind to both IL-6 itself and the two natural receptors have been produced.

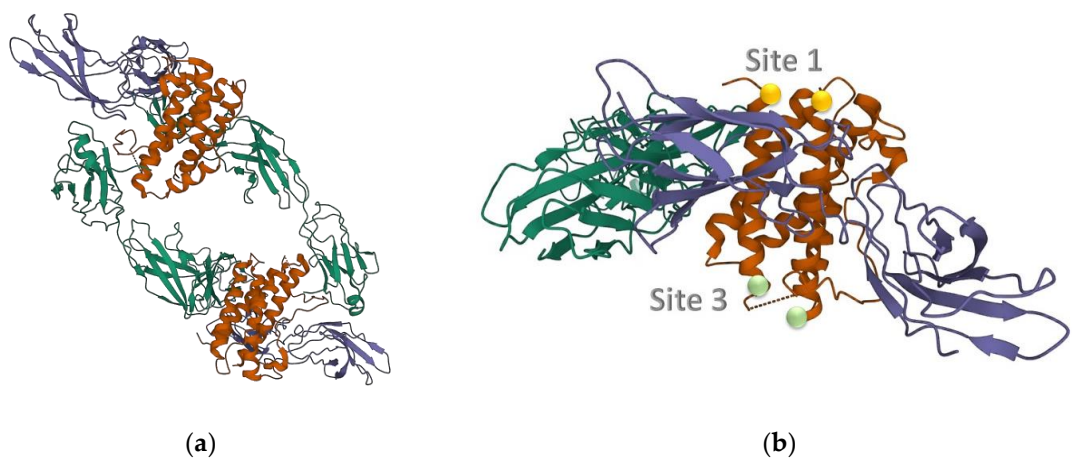

(**a**)  (**b**)

**Figure 1.** Quaternary structures of IL-6, cytokine IL-6 (orange) complexes with the receptor IL-6R, α-chain, (indigo) and the glycoprotein gp-130, β-chain (green). (**a**) Hexameric form; (**b**) Trimeric form; the contact positions named Site 1 and Site 3 are represented by small balls. (PDB entry: 1p9m) [38,43].

Various therapeutic MAbs (sarilumab and siltuximab, for instance) are currently used to inhibit IL-6, on the other hand, new macromolecules are continuously analyzed to specifically detect this cytokine. In this analysis, we compare the topological features of 5 receptors, quite different in structure and inhibition mechanisms: one mAb (olokizumab, OKZ) [39], still in the clinical trials, a pair of humanized camelid Fabs (61H7, 68F2) [44,45], and an aptamer modified to increase the binding time, a so-called SOMAmer (slow off-rate modified aptamer) (SL1025) [46,47]. All the proposed receptors have a high affinity with the cytokine, as reported in Table 1.

**Table 1.** IL-6 synthetic receptors analyzed in this paper. The affinity and the area shared with the cytokine, as resolved in the literature, are reported for each of them.

| Receptor | $K_D$ (pM) | Area * (Å) |
|---|---|---|
| OKZ | 10 [1] | 840 [2] |
| 61H7 | <10 [2] | 940 [2] |
| 68F2 | 10–20 [2] | 1160 [2] |
| SL1025 | 200 [3] | >1200 [3] |

[1] [39]; [2] [45]; [3] [46,47]; * *buried area*.

2.1.2. α-Thrombin Receptors

α-thrombin is a biomolecule fundamental for the formation of blood clots, which it initiates by catalyzing the cleavage of fibrinogen [48]. Its malfunction can produce, on the one hand, excessive thrombus formation and, on the other, bleeding phenomena.

The role of α-thrombin in the evolution of cancer has recently been highlighted [49–51]. Tumor cells induce an increase in the thrombin level both directly through the expression and release of procoagulant factors and indirectly through the activation of other cells, such as platelets, leukocytes, and erythrocytes. Furthermore, the use of chemotherapy improves the inflammatory state within a cancer condition and can stimulate cancer cells to be procoagulant. At least, high serum level of thrombin increases risk of thrombosis, but also promotes tumor growth and metastasis [51].

Some thrombin inhibitors have been developed for clinical applications: in addition to heparin and vitamin k, which are the best known, some other drugs have been synthetized and are commonly used in prophylaxis and treatment [52]. On the other side, most of the best-known inhibitors cause serious side effects.

The aptamers developed for thrombin, ATA, are a large and varied family, as they include several DNA and RNA aptamers, often chemically modified or in the chimera form. On thrombin, different binding sites are identified, and the main ones are indicated in [48] as exosite 1 (fibrinogen site) and exosite 2 (heparin site). The ATA chains are short, from 15 to 42 nucleotides, and flexible, thus allowing for an excellent adaptability to the ligand, as evidenced by the remarkable contact area [53] (Table 2). Specifically, we report data from two different teams, which are, otherwise, quite similar. On the other hand, data concerning the affinity of the cognate molecules are much more divergent and not a clear correlation with the common area that can be detected.

The aptamers analyzed in this work bind preferentially to exosite 1 or exosite 2, although some of them seem to favor a 1:2 stoichiometry, i.e., the same aptamer binds one protein to site 1 and one to site 2 (TBA, mTBA,) [48,54]. Most of the aptamers linked to exosite 1 are derived from TBA [55], a 15mer DNA oligomer from which they differ for specific modifications, such as lacking the thymine nucleobase at position 3, TBAΔT3, or at position 12, TBAΔT12 [56], or for a polarity inversion (mTBA) [57]; the substitution of thymine in position 4 with two different T-analogs [58] has produced high affinity aptamers (TBA-4W, TBA-4K). Finally, TBA itself was observed chelated with two different cations ($K^+$ and $NA^+$). Quite different from TBA are the 26mer NU172, chelated with $K^+$ and $Na^+$, [59] and the 31mer RE31 [60]. Furthermore, there is a sandwich form consisting of 2 aptamers, each bound to a different site (TerΔT3). Finally, two RNAs (Toggle-25t, mToggle-25t) [61] and one DNA(HD22) specific for site 2 are also analyzed [62].

**Table 2.** List of anti-thrombin aptamers analyzed in this paper. The interface area and $K_D$, as given in the literature, are reported.

| Receptor/PDB Code | $K_D$ ** (nM) | Area (Å) |
|---|---|---|
| TBA-$K^+$/4DII | 34 *** | 540 §/583 ** |
| TBA-$Na^+$/4DIH | 34 *** | 565 §/597 ** |
| mTBA/3QLP | 25 *** | 657 #/703 ** |
| TBAΔT3/4LZ4 | 55 *** | 524 §/548 ** |
| TBAΔT12/4LZ1 | 39 *** | 534 §/545 ** |
| TBA-T4W/6EO6 | 1 *** | 708 #/760 ** |
| TBA-T4K/6EO7 | 0.39 *** | 641 #/662 ** |
| TerΔT3_E/5EW1_E | 55 *** | 501 §§/531 ** |
| NU172-$K^+$/6EVV | 0.29 ** | 591 ##/656 ** |
| NU172-$Na^+$/6GN7 | 35 ** | 588 ##/643 ** |
| RE31/5CMX | 0.56 *** | 552 ##/661 ** |
| HD22/4I7Y | 29 *** | 1079 §§/1294 *** |
| TerΔT3_D/5EW1_D | 29 *** | 1116 §§/1135 *** |
| mToggle-25t/5DO4 | 0.002 *** | 925 #/1344 *** |
| Toggle-25t/3DD2 | 1.9 *** | 755 */1135 *** |

§ [53]; * [62]; ** [63]; *** [64]; # [65]; §§ [57]; ## [59].

*2.2. Methods*

Proteotronics [10] is the name we use to collect different kinds of analyses that share the idea of rendering the functioning of a macromolecule by means of an interaction network. We highlight that, hereafter, the term macromolecule will be used for the single object, ligand/receptor, and the ligand–receptor complex.

When the 3D structure is known, by means of crystallographic or in-silico techniques [12,13,66], a coarse-grained mapping is performed, collecting a set of 'nodes', one for each amino acid, each equipped with a set of data (position, electrical resistance, dielectric constant, etc.). Then, the nodes are connected among them, according to their distance, until the network is formed. The links represent "channels" for the exchange of information. When the information is encoded by means of an electric/electromagnetic signal, the network is put in contact with an external bias through ideal electrodes, and each link between nodes $j, k$ is associated to the impedance of a circuit element of area $A(j, k)$ and length $d(j, k)$, say $Z(j, k)$:

$$Z(j,k) = \frac{d(j,k)}{A(j,k)[\rho(j,l)^{-1} + i\varepsilon(j,l)\omega]}, \quad (1)$$

where $d(j, k)$ is the Euclidean distance between the nodes, $\rho(j, l)$, $\varepsilon(j, l)$ the resistivity and dielectric constant, respectively, and $\omega$ is the angular frequency of the external bias. The area: $A(j, k) = \frac{\pi(D^2 - d(j,k)^2)}{4}$ is defined for $D > d(j, k)$, with $D$ being the so-called cut-off distance, i.e., the maximal distance to define two connected nodes.

Finally, the complete macromolecule impedance can be calculated by solving a system of linear equations given by Kirchhoff's current law [10–17].

The cut-off value photographs the protein in a particular state of activation. In fact, the binding process, in addition to modifying the protein structure, also acts on the internal and external interactions, which become even more local [16]. In this perspective, varying $D$ is equivalent for describing the evolution of the protein.

Impedance (1) reduces to a simple resistance, $R(j, k)$, when $\omega = 0$, and for an assigned position of the contacts, it solely depends on the network topology. Therefore, one can imagine introducing some electrical charge inside the protein from one side and collecting it from another side: the measured resistance describes how long and impervious the charge path is, in other terms, to what extent the complexity of the macromolecule structure is. The use of the current flux as a topological probe is here enforced by assuming a single value for $\rho(j, k)$ [16]. Specifically, it can be used to analyze the interconnection between two cognate molecules.

Therefore, to perform this analysis, the resistance of the macromolecule is used, taken at different values of $D$. In particular, to have insights about the *topological complementarity* of different receptors of the same protein, we consider the ratio between the resistance of the ligand–receptor complex, $R_{comp\#}(D)$, and the resistance of the ligand, $R_{Lig}(D)$.

We remark that the *topological complementarity* is acquired during the binding process and is a tool to estimate how long the complex can survive. On the other hand, it is unlikely that the acquired *topological complementarity* also gives an estimate of the speed of the recognition-binding mechanism. In other terms, this quantity, as well as the interacting area, the buried surface, and the number of polar contacts, to mention the most known [63–65], cannot be used as an estimate of the affinity, which is a dynamical process depending on several factors, first of all, the ability of the cognate molecules to recognize and adapt to each other. Nevertheless, it can be used as a measure of the stability of the complex.

## 3. Results

### 3.1. Topological Investigation

A topological investigation is preparatory to further investigations because it provides some general insights into the elements of the macromolecules that are mainly involved in binding. It is based on the *contact maps* of the macromolecules, which are the graphical representation of adjacency matrices, i.e., Boolean matrices where 1 in the position *i,j* claims the presence of a link between the *i*-th and the *j*-th nodes, which, in turn, means that the corresponding nodes are closer than the cut-off distance, D. The contact maps are drawn for an assigned value of D, here D = 15 Å, which is the smallest value to completely resolve the macromolecules [10,13,15].

The 3D structures of the selected macromolecules are available in the protein data bank, PDB [43]. The PDB contains data coming from different teams, which can differ in some information, for example, the amino acid number. Therefore, to make our investigation reliable, here we select the strand GLU14-MET184 of IL-6 [38], which is common to all analyzed datasets. The contact map of the heterotrimer reveals a consistent amount of links between IL-6 and both natural receptors, IL-6R (mainly helices A,B,D) and gl-130 (mainly helices A,C), see Figure 2 and Table 3. We expect that the four helices involved in binding with natural receptors also play an active part in the interaction with different receptors. As a matter of fact, Fab 61HZ binds precisely to these helices (see Figures 2 and 3), while the aptamer (SOMAmer SL1025) binds only to three of them. OKZ implements a special kind of binding (see Figures 2 and 4), which, in addition to helices C,D, also involves helix E and forces the random coil F to organize itself into a small chain, thus blocking the coupling with gp-130. This form of steric hindrance seems to be the main one responsible for the inhibitory action of OKZ [39], which is very effective even at small doses, especially in the case of rheumatoid arthritis [39].

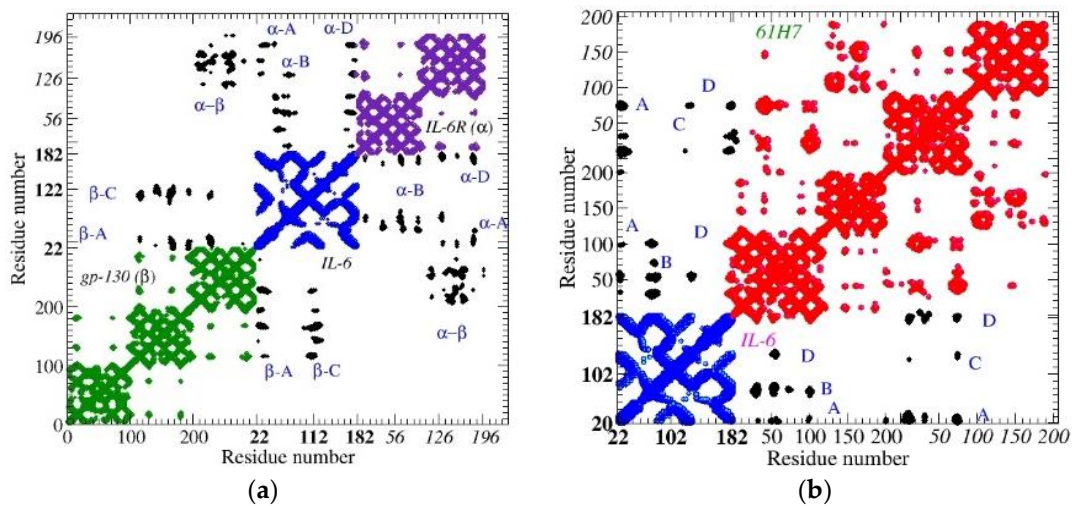

(a)    (b)

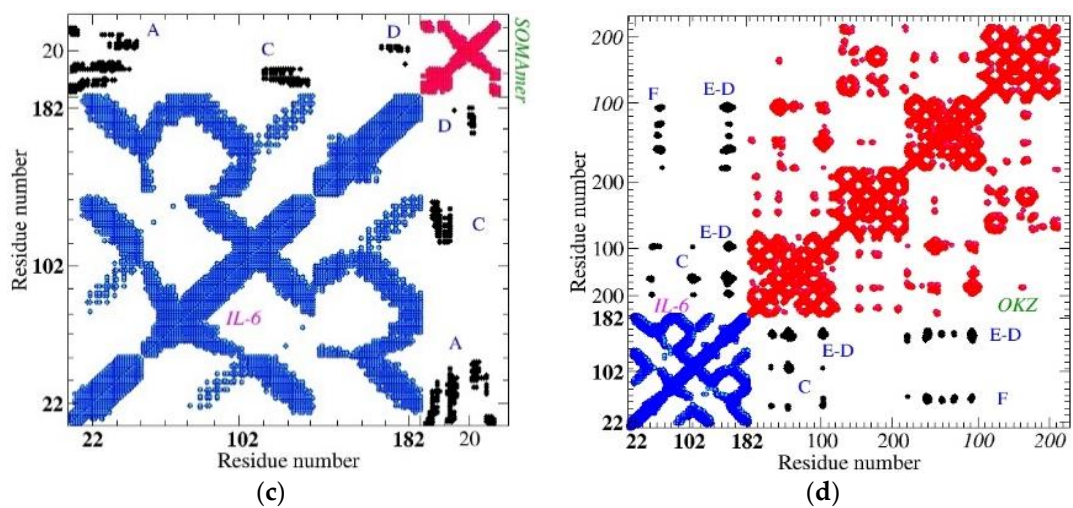

**Figure 2.** Contact maps of cytokine IL-6 complexed with different receptors. In all figures, cytokine IL-6 is drawn in blue and interlinks are in black. (**a**) the natural assembly gp-130*IL-6*IL-6R. Protein gp-130, i.e., receptor β, is drawn in green, protein IL-6R, i.e., receptor α, is drawn in indigo. The links between the receptors and the cytokine are drawn in black. The sequence of residues is reported on the axes, normal style for receptor β, bold for cytokine, slanted for receptor α (**b**) the assembly with humanized Fab 61H7 (in red); (**c**) the assembly with SOMAmer SL1025 (in red); (**d**) the assembly with olokizumab (in red). The cut-off value is 15 Å. Capital letters emphasize the main contacts with the cytokine helices, specifically, with helix A (SER21-SER47), helix B (GLU80-ARG104), C (SER107-LYS129), D (GLN156-ARG182), E (PRO141-GLN152).

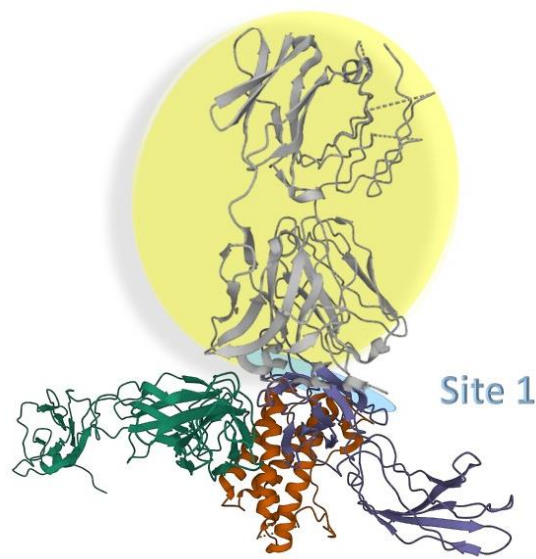

**Figure 3.** Schematic representation of the binding of Fab 61H7 (gray) to the cytokine. The picture is an artifact of the natural trimer and Fab 61H7 on site 1 and has the aim to make clearer the positions of different binding sites. Circles render site1 (cyan) and the antibody (yellow).

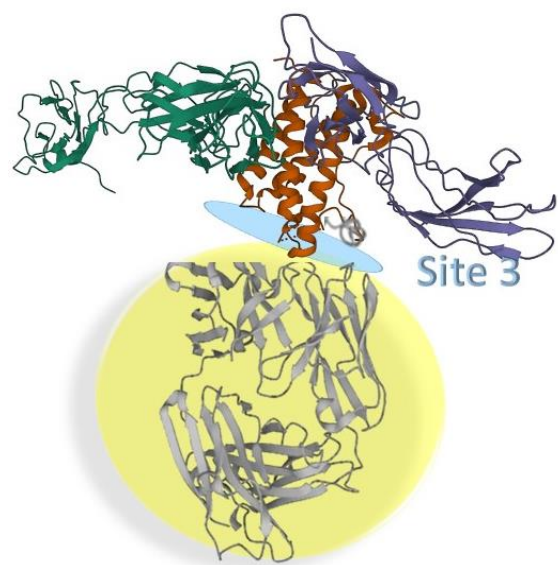

**Figure 4.** Schematic representation of the binding of OKZ (gray) to the cytokine. The picture is an artifact of the natural trimer with OKZ on site 3 and has the aim to make clearer the positions of different binding sites. Circles render Site 3 (cyan) and the antibody (yellow); the small helix E, up to Site 3 (gray) is pictured.

**Table 3.** List of IL-6 analyzed receptors. For each receptor, the IL-6 helices mainly involved in the binding process are reported, see Figure 2.

| Receptor | Helix |
|---|---|
| IL-6R | A, B, D |
| gp-130 | A, C |
| Fab 68F2 | A, D |
| Fab 61HZ | A, D, B, C |
| Antibody-OKZ | C, D, E, F |
| SOMAmer- SL1025 | A, C, D |

We highlight the high number of contacts observed between SOMAmer SL1025 and the cytokine (Figure 2c), and this is due to the great flexibility of this kind of receptor, which finely adapts to the ligand [46,47].

*3.2. Electronic Analysis*

We represent, as described in Section 2.2, the macromolecule (here the ligand) via an electrical network fed with two ideal electrodes placed near a binding site. In this way, we can follow the flow of charge inside it, both when it is isolated and when it is bound to the receptor.

In particular, the relative resistance, $rel_{res} = \frac{res_{comp\#}}{res_{Lig}}$, calculated at increasing values of D, provides various information on the complementarity between the two cognate molecules [16]. The relative resistance, $rel_{res}$, is calculated by placing the ideal electrodes close to the binding region, i.e., between the ligand and the receptor (Figures 3 and 4). This type of contacting puts the networks of the cognate molecules in parallel, with respect the electrodes and the resistance of the complex, which is always lower than that of the ligand. By increasing the cut-off value, D, the resistance of each macromolecule decreases. Furthermore, $rel_{res}$ also decreases, and the decrease rate depends on the complementarity between the two molecules: the better the interconnection, the lower the $rel_{res}$. The *rate* at which $rel_{res}$ decreases is an indicator of the *topological complementarity* of the cognate molecules. Furthermore, at the smallest values of D, if the two molecules are not close enough, the resistance of the complex coincides with that of the protein (on which the contacts reside), and for this reason, $rel_{res}$ assumes its maximum value. The *maximum* is close to 1 (the charge flows only through the ligand) when the cognate molecules are far apart. Finally, the $rel_{res}$ *asymptotic value* (large D) reflects the maximum amount of charge that can pass through the receptor, relative to that which passes through the protein: the better the adaptation and the smaller this value. Notice that the *asymptotic value* is affected by the receptor size, i.e., smaller receptors give larger *asymptotic values*.

Thus, for receptors of comparable size, such as the analyzed Fabs, gp-130, IL-6R, and OKZ, a small *asymptotic value* indicates a better fit. In the case of the aptamer, which is much smaller than the protein, the final value, which is rather high, is mainly due to the small size of the receptor, compared to the protein (most of the current passes through the protein) (Figure 5). To account for all this information, in particular, the (i) value of the maximum, (ii) rate of decrease, and (iii) *asymptotic value*, we calculate the ToCI (topological complementarity index) indicator, which is obtained by multiplying the area subtended by $rel_{res}$ curve when plotted as a function of D, normalized to the number of nodes (amino acids + nucleobases) in the complex [16]. This quantity depends on the position of the electrodes and is minimal when the electrodes are on the binding region.

Smaller values of ToCI suggest a better complementarity.

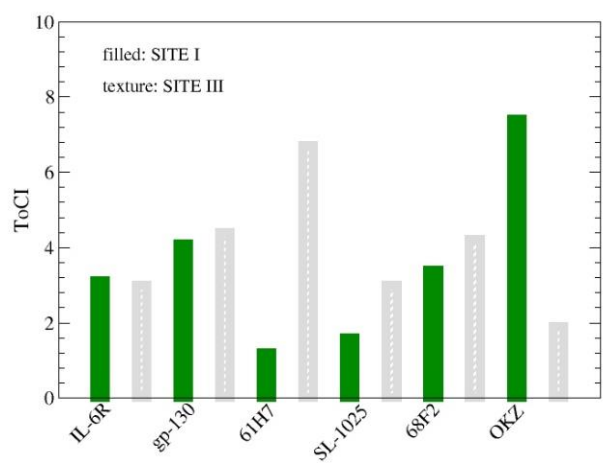

**Figure 5.** ToCI for the analyzed receptors of cytokine IL-6. The contacts are put on Site I (green), which is the natural binding site for all the listed receptors but olokizumab (OKZ), which has Site III as the natural binding site. Largest differences between the two Sites are observed for the humanized Fab 61H7, while for the natural receptors (IL-6R, gp-130), which are arranged symmetrically, with respect to the two sites (see Figures 3 and 4), the differences are tiny.

3.2.1. ToCI for Cytokine IL-6 Receptors

ToCI was calculated for the different positions of the electrodes. In particular, two different locations were considered, the first, indicated as Site 1, with the contacts on GLU23 and MET184 [38], and the second, indicated as Site 3, with the contacts on (THR23 and ALA130) (Figures 3 and 4). The ToCI value reduces, in some cases even significantly (e.g., OKZ [38,67], Fab61H7 [45]), when the electrodes are on the region of greatest contact. The relative resistances were calculated for the receptors listed in Table 3.

Figure 5 reports the bar chart of the values of ToCI, calculated using two different electrode positions. The best results are for the Fab 61H7 and SOMAmer SL1025 (SITE I) and OKZ (SITE III), in agreement with the regions of maximal binding. IL-6R and gl-130 give similar results by using the two different contact positions: this is due to their axial symmetric distribution around the cytokine (Figure 1). Notice that the small value of ToCI for the aptamer is in agreement with its specific ability to remain stably bound to the target.

A significant correlation between ToCI and the affinity or common area reported in the literature cannot be tested, due to the reduced size of the dataset: this kind of analysis will be performed with the receptors of a-thrombin.

3.2.2. ToCI for α-Thrombin Receptors

Following the procedure seen in the previous section, the electrodes were positioned close to the main binding sites (Section 2.1.2). Specifically, the first pair of electrodes on the amino acids GLU39, GLU80, [68] (SITE 1), while the second, on ASP125, ASP178 [56,68] (SITE 2).

The two contact locations give quite different ToCI values, meaning that the aptamers extend close to only one of the sites.

In Figure 6, we report the ToCI values of two groups of aptamers, those that bind to SITE I (left) and those that bind to SITE 2 (right). Regarding the first group, the best complementarity between α-thrombin and aptamers is obtained by TBA-T4W, which also shows the largest shared area of this group (Table 2). As for the aptamers that bind to SITE 2, all outperformed the first group. Note that the area that the aptamers of the second group share with the protein is about twice that shared by the aptamers of the first group.

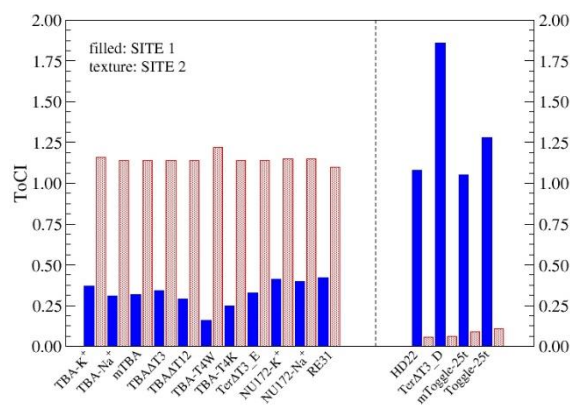

**Figure 6.** ToCI for the analyzed receptors of α-thrombin. The contacts are put on SITE I (blue) and SITE II (red). Aptamers HD22, terΔT3_D, Toggle-25t, and mToggle-25t bind to SITE 2. Toggle-25t and mToggle-25t are RNA aptamers.

Finally, no significant differences between the RNA aptamers (Toggle-25t and mToggle-25t) and the other aptamers of the second group are highlighted by this procedure

As discussed in Section 2.2, a clear correlation between the area shared by the complex (buried area or interface area) and $K_D$ [64] cannot be identified. On the other hand, a substantial correlation between the common area and ToCI is detected in Figure 7, which represents the scatter plot of ToCI and the interface area. In that figure, we report data from two different teams, as a matter of fact, for both datasets, the correlation coefficient is high (close to −0.9), which confirms ToCI as a good tool for evaluating the complementarity of the cognate molecules.

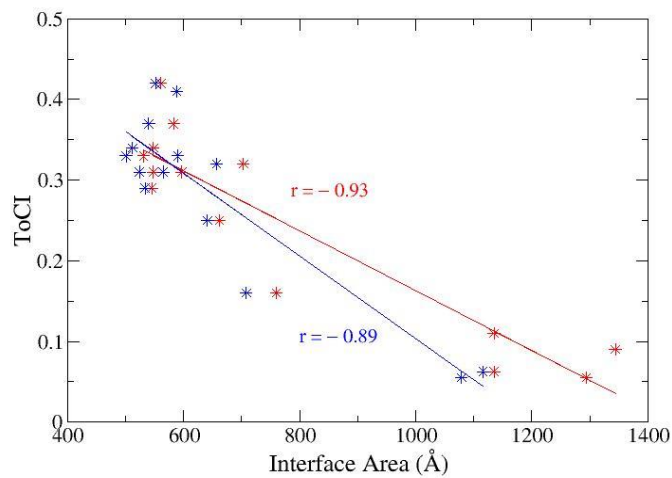

**Figure 7.** Scatter plot of the ToCI vs shared area as measured by different teams **red** (interface area) [64], **blue** (buried area) [59,65]. Stars represent data, lines the linear regression fittings, r is the correlation value.

## 4. Discussion

In this paper, we have described a strategy to analyze the conformational complementarity of receptor–ligand complexes. Conformational complementarity is an interesting tool to evaluate the stability of the complex, which, in turn, is an important parameter to consider for the design of biodevices and of drugs and drug delivery systems.

We have analyzed the complexes obtained from biomolecules capable of modifying their structure to perform an optimal adaptation: this is what normally happens when, for example, a protein receptor recognizes and binds to its specific ligand. On the other hand, in biodevices, mutual adaptation is not the only mechanism that can be implemented. Indeed, the target can also be captured in its native form, using, for example, MIPs, which do not implement forms of conformational adaptation. In this case, the stability can be even higher than in the case of biomaterials that implement the adaptation, although the capture mechanism can be less efficient.

Stability analysis focused on aptamers and Fabs implementing a binding mechanism based on mutual adaptability. The strategy we follow considers the pair of biomolecules as a network formed by two interconnected impedance sub-networks, in which an electric current can flow. Charge flow is enabled by a pair of ideal contacts strategically placed on the ligand binding site. The observed flux is greater for the complex than for the single ligand, and the better the complementarity of the biomolecules, the higher the flux. The flow value is analyzed by gradually increasing the range of distances at which the complex is analysed. At the smallest distances, only complexes composed of very close biomolecules allow for charge flow, and all the charges flow inside the ligand (only one of the subnetworks is active). As the distance increases, the subnets become even more connected, their effective resistance decreases, and the intra-complex flux becomes even smaller than the intra-complex flux. The integral of the flux over the distance interval, suitably normalized to the size of the complex, provides the indicator of complementarity.

The analysis carried out on some $\alpha$-thrombin receptors confirms the experimental information that, for the SL1025 aptamer, whose stability is notoriously high, the ToCI is rather low, as well as for the antigen-binding fragment of clinical interest, OKZ. Furthermore, an interesting value was obtained for the camelid Fab 61H7.

Finally, some IL-6 cytokine receptors were analyzed, in this case, all aptamers. They are divided, both by size and by the site to which they are linked, into two groups, with very different characteristic values of the ToCI. The attamers that bind to SITE 1 show higher ToCI values than those that bind to SITE 2; furthermore, the minimum value of this group belongs to the experimentally best performing one (TBA-T4W).

## 5. Conclusions

Biologically inspired materials are increasingly becoming part of electronic devices. The reason is that they are more easily engineered and more customizable than their inorganic analogs, are nano-sized, and mimic, if not improve on, the functions already made very efficient by millions of years of evolution. The choice of a specific biomaterial depends on its intended use, for example, in the case of sensors, on the type of expected response, and on the use of the device (whether used on a large scale or in the laboratory). When choosing, various variables must be considered, from the ease of availability of the biomaterial to the selectivity and sensitivity and stability of the device to be constructed.

Some biomolecules, such as aptamers and antibodies, can give excellent results, in terms of selectivity and stability of the response. In fact, they produce a stable bond with the target, thanks to the innate flexibility that allows them to adapt to this, when they are in their activated state. In this way, the target molecule can be recognized when it is already functional. Furthermore, this same type of material can also respond to non-material stimuli, i.e., light, allowing for the creation of a new generation of photovoltaic cells [10].

On the other hand, although the study and development of this type of material has been going on for a long time [20], some limitations of use, i.e., the high costs and the long production times, in the first place, can limit its large-scale diffusion.

Alternatively, the use of polymers in the fabrication of MIPs is quite cheap and rapid and, therefore, potentially allows for large-scale use [29]. The application limit of this technology is that it does not provide for the adaptation of either the polymer or the target. Therefore, this type of procedure selects the non-activated fraction of the target molecules and can be very interesting for a high selectivity analysis [30,31]. On the other hand, the activation of the target molecule, with the consequent change of structure, reduces the efficiency of the detection performed with this technique.

Finally, the characteristics of reliability, selectivity, and adaptation to the target are optimized by integrating extremely flexible and adaptable macromolecules, such as aptamers in stable and reliable platforms, such as MIPs [69]: integration is also, in this case, a trump card.

**Author Contributions:** Conceptualization and formal analysis, E.A.; investigation, E.A., M.C., A.G.M., G.M., and S.R.; writing—original draft preparation, E.A.; writing—review and editing, A.G.M., G.M., and S.R. All authors have read and agreed to the published version of the manuscript.

**Funding:** This research received no external funding.

**Informed Consent Statement:** not applicable

**Data Availability Statement:** The data presented in this study are available on request from the corresponding author.

**Acknowledgments:** E.A. acknowledges R. Cataldo (University of Salento) for preliminary discussions about some topics covered in this review.

**Conflicts of Interest:** The authors declare no conflict of interest.